\documentclass{PoS}

\newcommand{\fet}[1]{\mbox{\boldmath $#1$}}

\title{Nuclear Physics with Chiral Effective Field Theory: State of the Art and Open Challenges}

\ShortTitle{Nuclear Physics with Chiral Effective Field Theory}

\author{\speaker{Evgeny Epelbaum}\thanks{It is a great pleasure to
    thank all my collaborators for sharing their insights into the 
discussed topics and the organizers of Confinement X
for making this exciting conference possible. I also thank Ashot
Gasparyan, Jambul Gegelia, Hermann Krebs, Timo L\"ahde, Deen Lee and
Ulf-G.~Mei{\ss}ner for reading the manuscript and making many valuable
suggestions. Finally,  I acknowledge
financial support by the 
European Research Council (ERC-2010-StG 259218 NuclearEFT) and the
Deutsche Forschungsgemeinschaft (SFB/TR 16). }\\
        Institut f\"ur Theoretische Physik II, Ruhr-Universit\"at Bochum,
  D-44780 Bochum, Germany\\
        E-mail: \email{evgeny.epelbaum@rub.de}}

\abstract{Understanding the properties of atomic nuclei and nuclear dynamics
from QCD remains a major challenge. Complementary to first
attempts along these lines based on lattice QCD, an effective field
theory  approach has been developed in the past two decades and applied to a variety of
nuclear bound states and reactions. I outline the foundations of
this method, review selected applications and address some
 open challenges in this field.}

\FullConference{Xth Quark Confinement and the Hadron Spectrum,\\
		October 8-12, 2012\\
		TUM Campus Garching, Munich, Germany}

\begin{document}

\section{Introduction}

The past decade has witnessed a renewed interest in the nuclear
force problem. In addition to new experimental facilities, this is to
a large extent related to exciting theoretical developments in
this field. On
the one hand, rapidly increasing computational resources and
improvements in algorithms make some nuclear physics observables
amenable to numerical simulations in lattice QCD \cite{Detmold}. 
Complementary to this research direction, considerable
progress has been achieved towards a quantitative description of
nuclear forces and dynamics within the framework of effective field
theory (EFT) starting from the pioneering work of Weinberg
\cite{Weinberg:1990rz}. The essential idea behind this method is to
systematically exploit the scale separation in a problem of
interest. There are several scales that play an important role in
nuclear physics. The lowest one is given by the typical binding
energies of the order of a few MeV per nucleon. This small scale
manifests itself in the large values of the nucleon-nucleon (NN) S-wave scattering
lengths, 
$a_{1S0} \sim (8\mbox{ MeV})^{-1}$ and $a_{3S1} \sim (40\mbox{ MeV})^{-1}$, 
and signals the breakdown of perturbation theory for two- and
many-nucleon observables at low energy.  It is well separated from the
next-higher scale relevant for the NN system, namely the pion
mass $M_\pi$. Exploiting this scale separation allows one to set up the
so-called pionless EFT. In this approach,
(non-relativistic) nucleons are treated as the only active degrees of
freedom (DOFs) and the
shallow scale associated with the nuclear binding is 
generated dynamically via resumming the lowest-order NN
interactions. Pionless EFT is justified for momenta well below
$M_\pi$ which is sufficient for e.g.~many reactions of astrophysical
interest. Similar theoretical methods are successfully  applied to study
Efimov physics and  universality in few-body systems close to the unitary
limit, cold atoms and the properties of halo-nuclei, see
e.g.~\cite{Hammer:2010kp} for a recent
review article.  

To increase the applicability range of the theory beyond the
near-threshold region one needs to include pions
as explicit DOFs. The
resulting chiral EFT ($\chi$EFT) relies on the approximate spontaneously broken
chiral symmetry of QCD. This symmetry/symmetry-breaking pattern of QCD
strongly constrains the interactions of pions being identified with the corresponding
pseudo-Goldstone bosons. It allows one to calculate pion and 
pion-nucleon low-energy observables within a systematic perturbative
expansion in $Q / \Lambda_\chi$. Here, $Q\sim M_\pi$ refers to the soft scale
associated with external momenta while $\Lambda_\chi \sim
M_\rho$ GeV stands for the (hard) chiral-symmetry breaking scale that 
governs the values of renormalized low-energy constants (LECs) in the
effective Lagrangian.  We refer the reader to the review
article \cite{Bernard:2007zu} and references therein for more details on chiral
perturbation theory (ChPT) and an overview of recent trends and developments
in that field. 

\begin{figure}[t!]
\begin{center}
\includegraphics[width=0.48\textwidth,keepaspectratio,angle=0,clip]{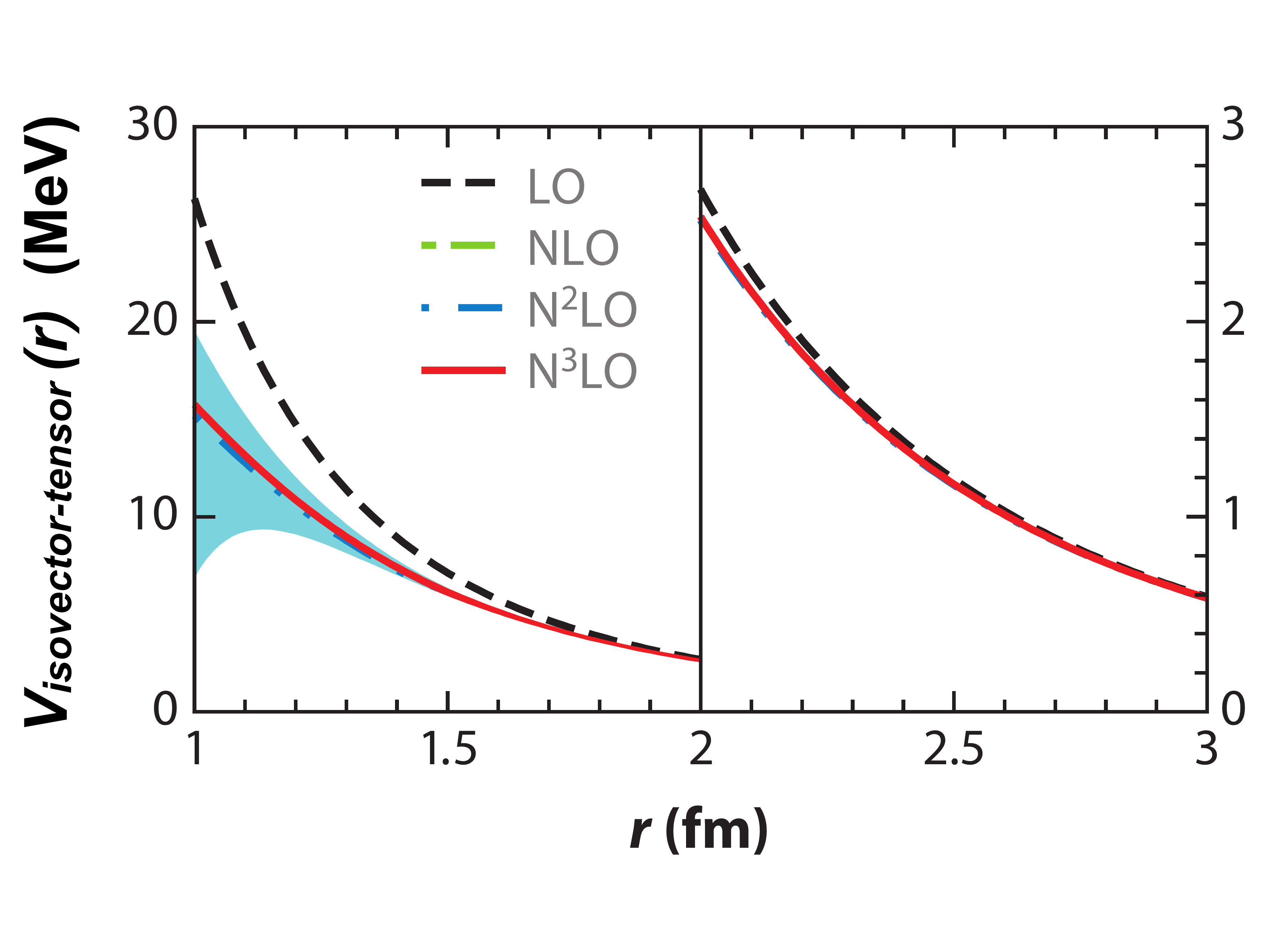}
\hfill
\includegraphics[width=0.48\textwidth,keepaspectratio,angle=0,clip]{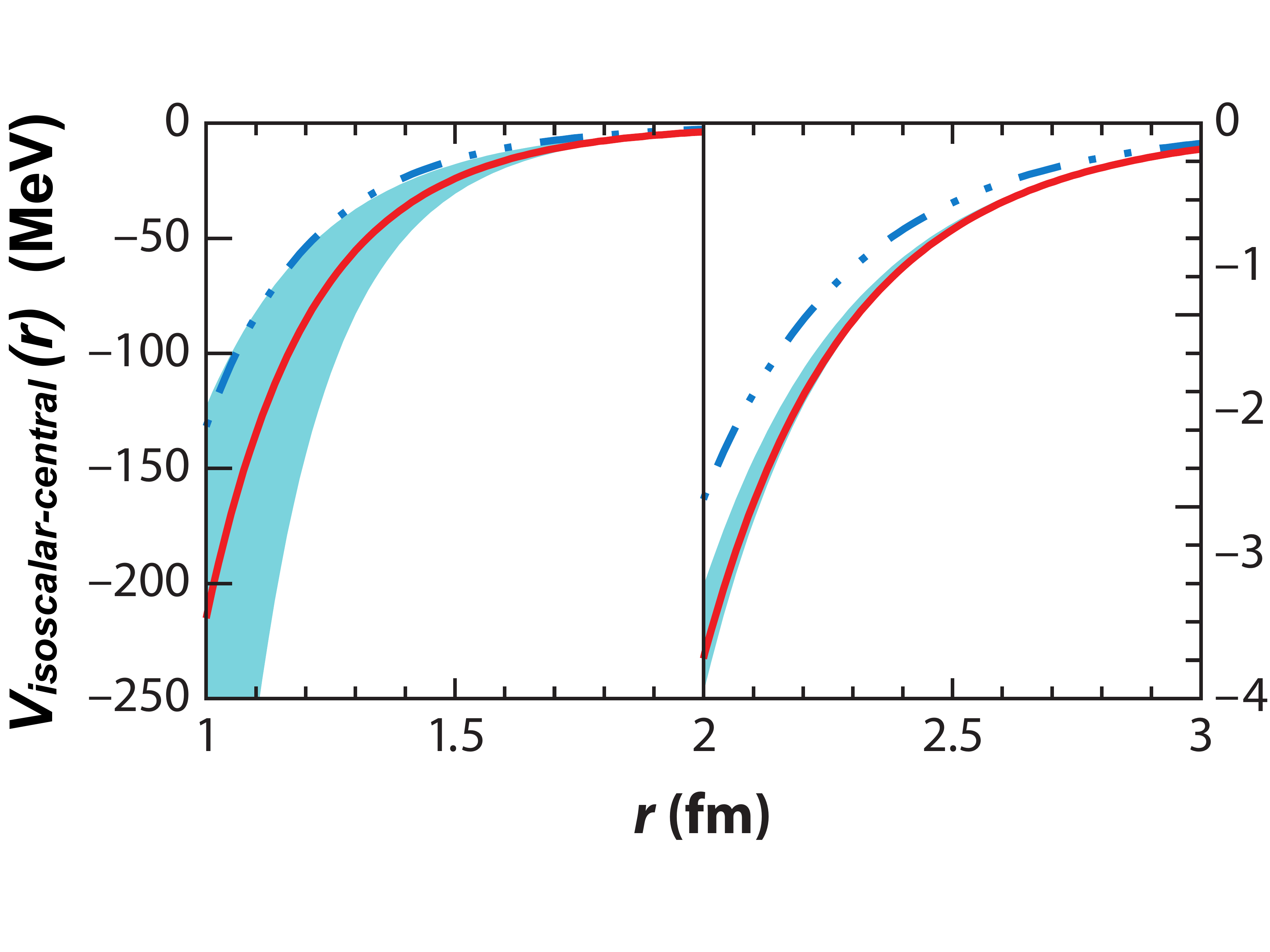}
\end{center}
\vspace{-1cm} 
    \caption{
         Chiral expansion of the isovector-tensor (left panel) and
         isoscalar central (right panel) long-range two-nucleon potentials. The shaded bands show
         an estimated size of (scheme-dependent) short-range
         contributions which are represented by contact interactions, see
         Ref.~\cite{Epelbaum:2012vx} for more details. 
\label{fig1} 
 }
\end{figure}

In the past decades, $\chi$EFT has been extensively applied to the
nuclear force problem, see \cite{Epelbaum:2012vx} for an 
introduction and \cite{Epelbaum:2008ga,Machleidt:2011zz} for recent
review articles. In this framework, nucleons interact 
by exchanging a single or multiple pions.  In the chiral limit of
 vanishing quark masses one is expanding about, these interactions 
 have an infinitely long range. The long-range tail of the
 nuclear force controls the energy dependence of the scattering
 amplitude.  It is strongly constrained by the chiral symmetry of QCD
 and can be rigorously derived in ChPT, see Fig.~\ref{fig1}. 
On the other hand, the short-range part of the range well below $M_\pi^{-1}$ 
 is driven by physics that cannot be resolved explicitly in
 reactions with typical nucleon momenta of $\mathcal{O}( M_\pi )$. Such short-range forces can be naturally  parameterized by
contact  interactions with an increasing number
 of derivatives. 

In this contribution I review the current status of $\chi$EFT for
nuclear forces and few-nuclear systems, discuss some recent
and ongoing developments and address open questions and
challenges in this field. The paper is organized as follows. Section
\ref{sec:forces} is devoted to nuclear forces and nuclear chiral 
dynamics. A few recent applications to few-nucleon reactions
with external probes are discussed in section 
\ref{sec:applications}. Finally, section \ref{sec:lattice} overviews 
progress towards understanding the properties of light nuclei via 
lattice simulations of $\chi$EFT. 

\section{Chiral effective field theory for nuclear forces and light nuclei}
\label{sec:forces}

Assuming that LECs accompanying few-nucleon contact interactions scale according to
naive dimensional analysis, the chiral power counting provides a
natural qualitative explanation of the (always assumed) dominance of
the two-body interactions with $\langle V_{\rm 2N} \rangle \gg 
\langle V_{\rm 3N} \rangle \gg \langle V_{\rm 4N} \rangle \gg
\ldots$. More precisely, the chiral expansion of nuclear forces has
the form \cite{Weinberg:1990rz}
\begin{eqnarray}
\label{nuclforces}
V_{\rm 2N} &=& V_{\rm 2N}^{(0)}+ V_{\rm 2N}^{(2)}+ V_{\rm 2N}^{(3)}+
V_{\rm 2N}^{(4)} + \ldots , \nonumber \\
V_{\rm 3N} &=&  V_{\rm 3N}^{(3)}+
V_{\rm 3N}^{(4)} + \ldots , \nonumber \\
V_{\rm 4N} &=&  V_{\rm 4N}^{(4)}+ \ldots , 
\end{eqnarray}
where the superscripts denote the associated powers of the soft scale $Q$. 

\subsection{The two-nucleon system}
\label{sec:NN}

Starting from the pioneering work by Weinberg \cite{Weinberg:1990rz}  
and the first quantitative calculation of Ref.~\cite{Ordonez:1993tn}, 
the NN force has been extensively studied in the framework
of $\chi$EFT. Within the heavy-baryon formulation,
calculations have been pushed to leading
two-loop order corresponding to next-to-next-to-next-to-leading
order (N$^3$LO) or $Q^4$ in the chiral expansion. At this order, the long-range part
of the NN force is governed by exchange of up to three pions. It is
strongly constrained by the chiral symmetry of QCD and experimental data on
pion-nucleon scattering. The short-range part depends on $26$ LECs accompanying
$24$ isospin-invariant and $2$ isospin-breaking NN contact
interactions which are tuned to the low-energy NN data
\cite{Entem:2003ft,Epelbaum:2004fk}.    
It was found to be necessary and
sufficient to go to N$^3$LO in order to accurately describe NN 
phase shifts up to energies of the order of $E_{\rm lab} \sim 200$ MeV. 
\begin{figure}
\centerline{\includegraphics[width=0.48\textwidth,keepaspectratio,angle=0,clip]{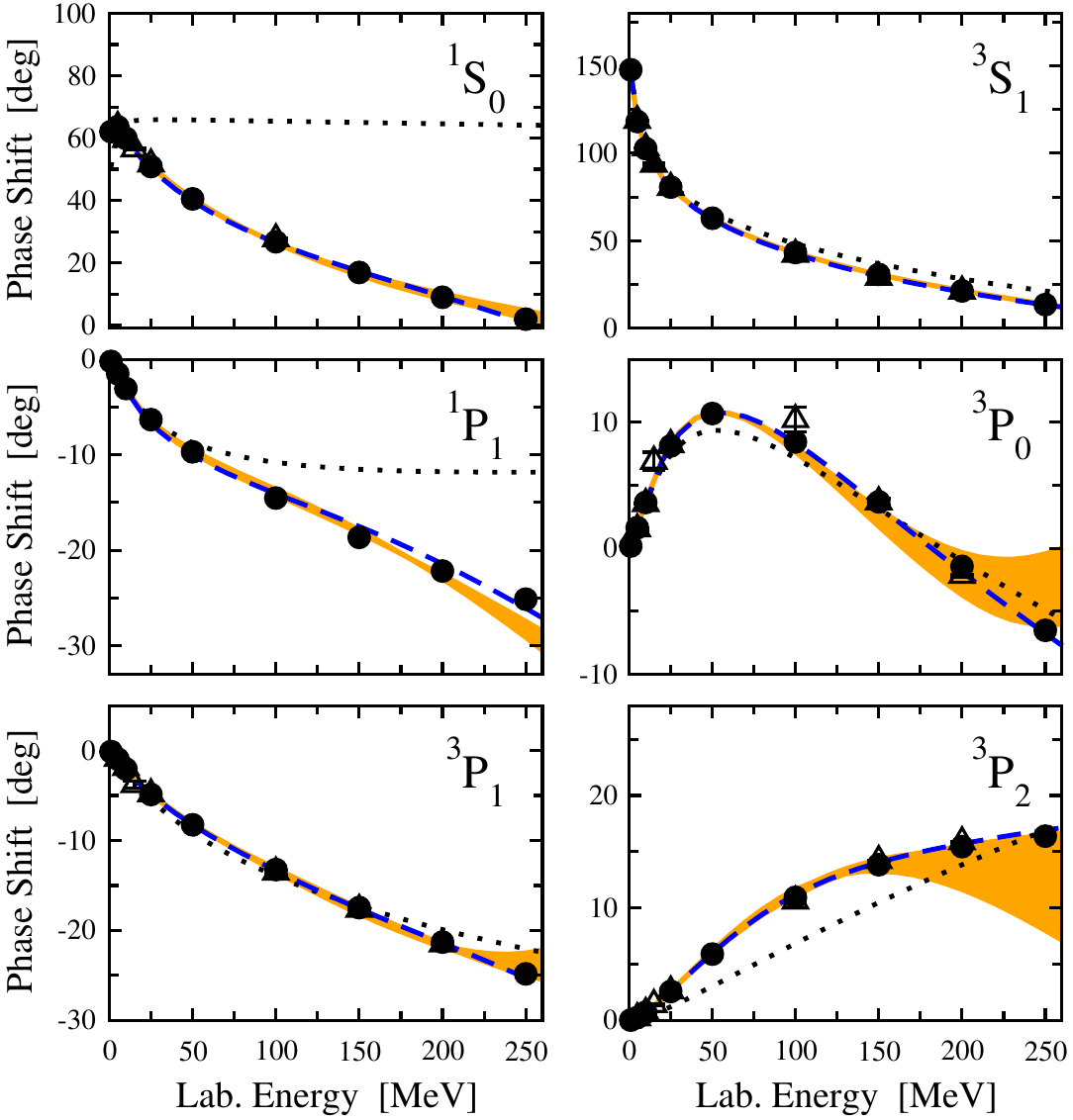}
\hfill
\includegraphics[width=0.48\textwidth,keepaspectratio,angle=0,clip]{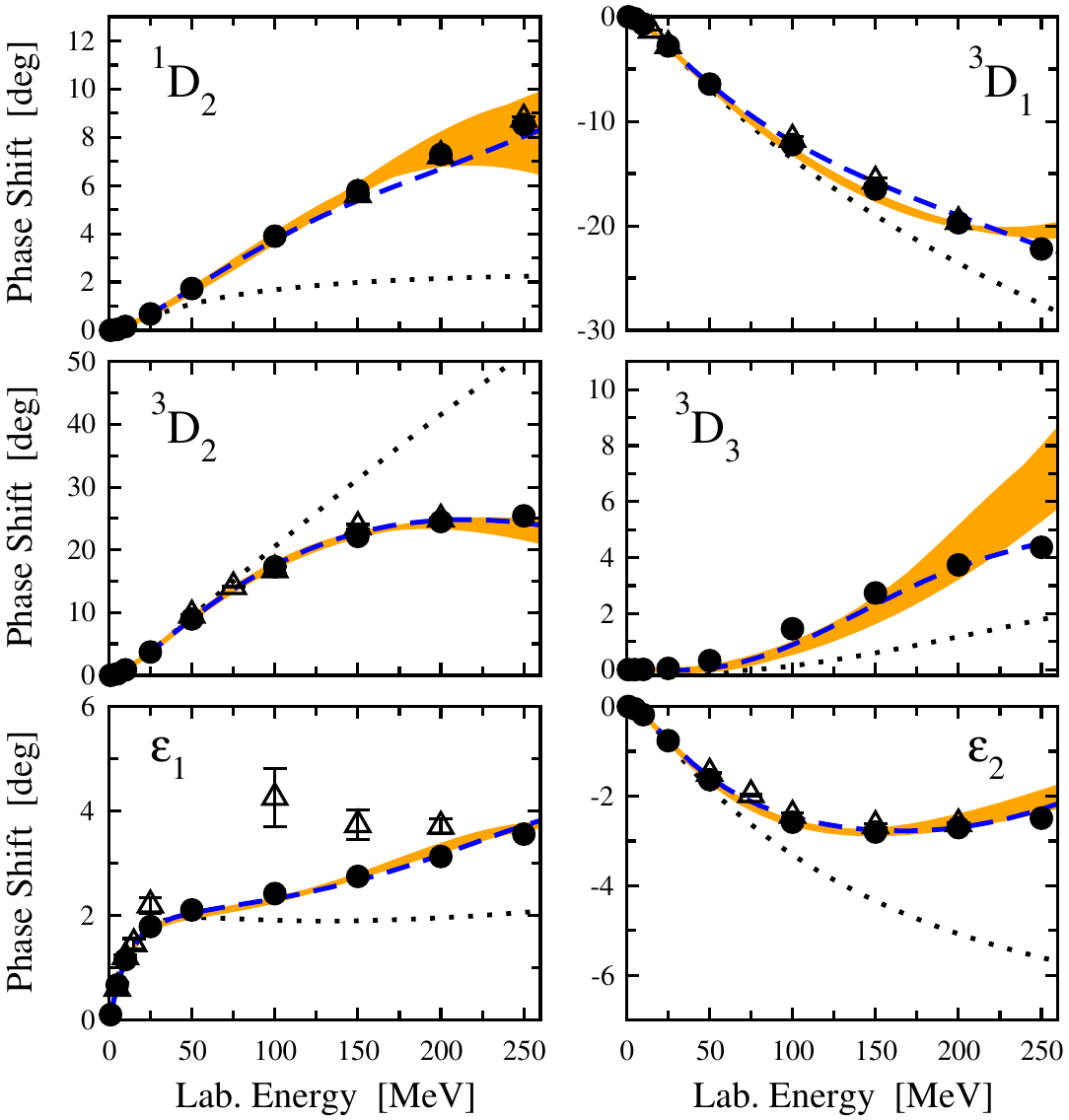}}
\caption[]{Neutron-proton phase shifts and mixing
  angles calculated using N$^3$LO $\chi$EFT potentials of Ref.~\cite{Epelbaum:2004fk}
 (shaded bands) and Ref.~\cite{Entem:2003ft} (dashed lines) in  
  comparison with the Nijmegen \cite{Stoks:1993tb} (filled circles) and 
  SAID \cite{SAID} (open triangles) partial wave analyses. Also shown
  are leading-order cutoff-independent results of Ref.~\cite{Epelbaum:2012ua} (dotted
  lines).  }
\label{fig:phases}
\end{figure}

The most interesting part of the novel chiral NN force is
two-pion ($2\pi$-) exchange which constitutes the second-longest
contribution to the NN potential and, therefore, has significant
impact on the energy dependence of the scattering amplitude. 
Indeed, its evidence has been confirmed in
the partial wave analysis of the Nijmegen group
\cite{Rentmeester:2003mf}, see also \cite{Birse:2003nz}. 
In agreement
with expectations based on phenomenological studies, one
observes a very strong attractive isoscalar central potential. This by
far the strongest $2\pi$-exchange contribution emerges, however, only at
next-to-next-to-leading order (N$^2$LO) as a correction to the
nominally dominant $2\pi$-exchange potential at next-to-leading
order (NLO). This peculiar pattern is well
understood and 
can be traced back to
the intermediate excitation of
the $\Delta$(1232) isobar at one of the nucleons which gives rise to a
very strong attractive isoscalar central NN force \cite{Ordonez:1993tn,Kaiser:1998wa,Krebs:2007rh}. 
In the standard formulation of $\chi$EFT
based on pions and nucleons as the only explicit DOFs,
all effects of the $\Delta$ (and heavier resonances as well as heavy mesons) are hidden
in the (renormalized) values of (some of the) LECs starting from the
subleading effective Lagrangian. As a consequence, the
phenomenologically important $2\pi$-exchange mechanism driven by
the $\Delta$ excitation appears only at subleading order from diagrams
involving one insertion of the subleading pion-nucleon vertex. The
values of the corresponding LECs $c_{3,4}$ are, to a large extent, driven by the $\Delta$
isobar \cite{Bernard:1996gq} and turn out to be rather large in magnitude. 
It is possible to improve the convergence of the  EFT expansion by
treating the $\Delta$-isobar as an explicit DOF in the
effective Lagrangian and counting $m_\Delta - m_N \sim M_\pi = {\cal
  O} (Q)$ \cite{Hemmert:1997ye}, see also \cite{Pascalutsa:2002pi} for an alternative counting scheme.   
In such a $\Delta$-full theory, the major part of the strong attractive
$2\pi$-exchange potential is shifted from N$^2$LO to NLO, while the  
LECs $c_{3,4}$ take more natural values \cite{Krebs:2007rh}.  

\begin{figure}
\centerline{\includegraphics[width=0.293\textwidth,keepaspectratio,angle=0,clip]{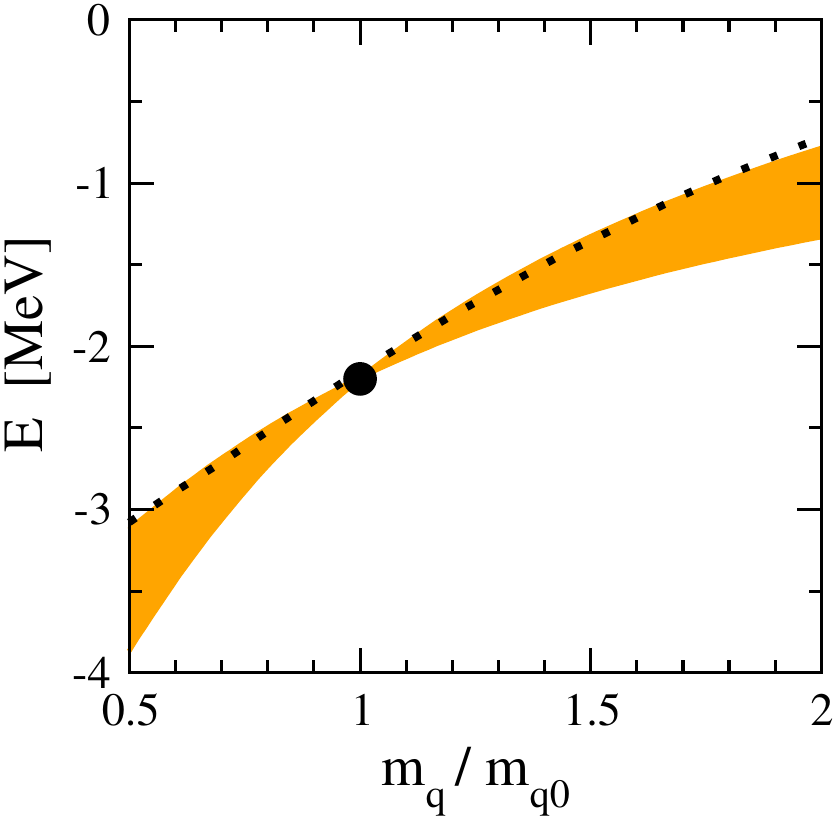}
\hfill
\includegraphics[width=0.65\textwidth,keepaspectratio,angle=0,clip]{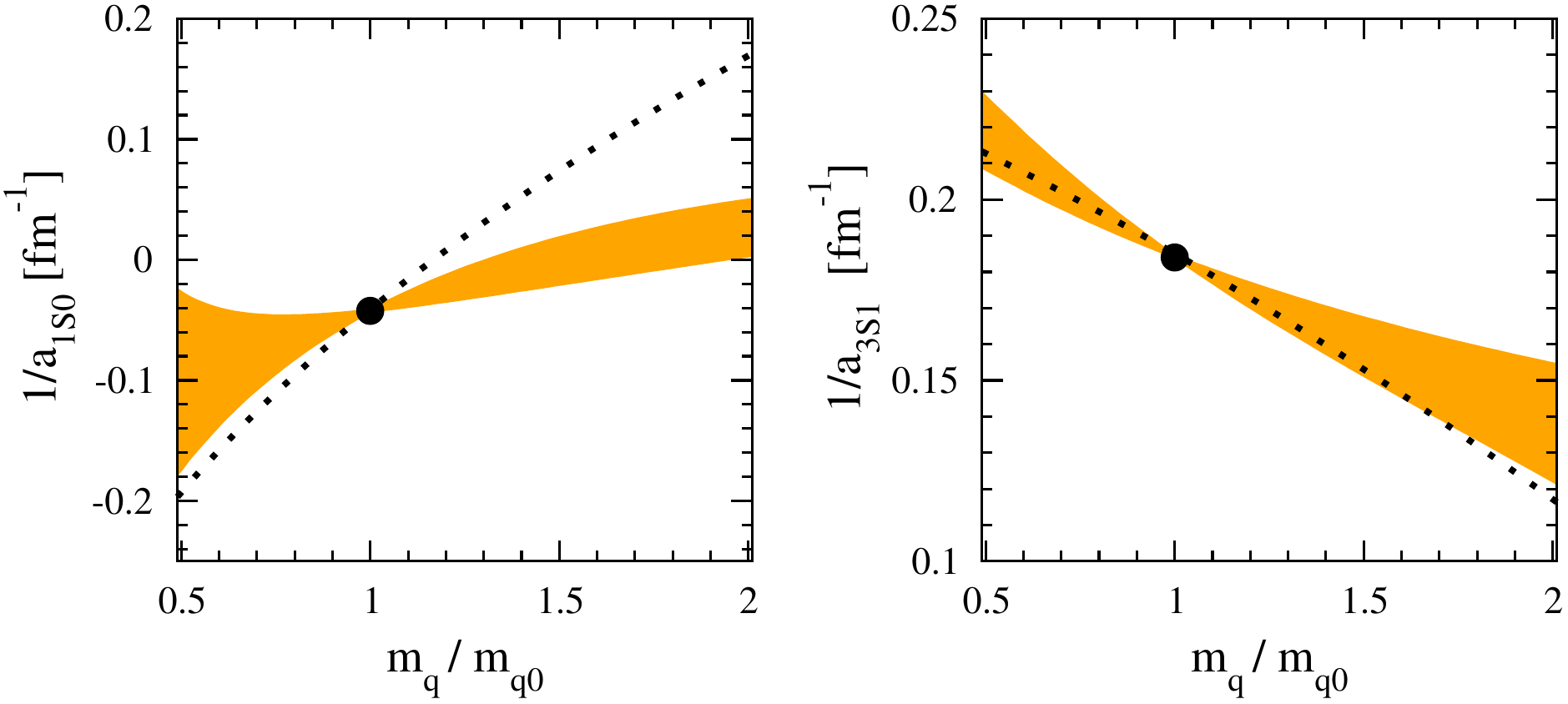}}
\caption[]{\label{fig:mpidep} Quark mass dependence of the deuteron
  binding energy (left panel), and inverse $^1S_0$/$^3S_1$
  neutron-proton scattering lengths (middle/right pannel). Shaded bands
  correspond to the N$^2$LO analysis of Ref.~\cite{Berengut:2013nh} as explained in
  the text. Also shown are leading-order cutoff-independent results of
Refs.~\cite{Epelbaum:2012ua,Epelbaum:2013ij}.}
\end{figure}

Having developed $\chi$EFT for the NN system, it is natural to address
the question of the light quark-mass- ($m_q$-) dependence of the nuclear force and
observables such as e.g.~the deuteron binding energy $E_{^2\rm H}$ and S-wave
scattering lengths $a_{1S0}$, $a_{3S1}$. This is not only of considerable interest for
ongoing and upcoming lattice-QCD calculations, but also for
searches of a possible spatial and temporal variation of fundamental
constants in nature \cite{Bedaque:2010hr} and questions related to anthropic considerations,
see also section \ref{sec:lattice}.  The $m_q$-dependence of NN S-wave
phase shifts and $E_{^2\rm H}$ was analyzed at NLO in
Ref.~\cite{Epelbaum:2002gb}, see also Ref.~\cite{Beane:2002xf} for a calculation using the
power counting scheme of Ref.~\cite{Kaplan:1998we}, which relies on a perturbative treatment of
$1\pi$-exchange, and more recent related studies \cite{Chen:2010yt,Soto:2011tb}.     
The common problem in all these calculations is the lack of
knowledge about the $m_q$-dependence of NN contact
interactions. Estimating the size of the corresponding LECs 
by means of
dimensional analysis leads to a very large uncertainty for chiral
extrapolations of   $E_{\rm ^2H}$, $a_{1S0}$ and $a_{3S1}$. In
addition, there are indications that the chiral expansion of 
the short-range part of the NN force might
converge slowly in the heavy-baryon approach due to the appearance of the
momentum scale $\sqrt{M_\pi m_N}$ associated with radiative pions \cite{Mondejar:2006yu}. 
To overcome these difficulties, the recent N$^2$LO analysis of Ref.~\cite{Berengut:2013nh}
made use of the fact that the LECs accompanying NN contact
interactions are saturated by heavy-meson exchanges
\cite{Epelbaum:2001fm,Epelbaum:2005pn}.  
Using a unitarized version of ChPT in combination
with lattice-QCD results to describe the $m_q$-dependence of meson
resonances saturating these LECs, the $m_q$-dependence of NN
observables was analyzed at N$^2$LO without 
relying on the chiral expansion of the short-range NN 
force, see Fig.~\ref{fig:mpidep}.  
This allowed us to considerably reduce the theoretical uncertainty as
compared to the earlier calculations. Extending these results to light
nuclei and comparing observed and calculated primordial deuterium and
helium abundances yields a stringent limit on a variation of the
light quark mass, $\delta m_q /m_q = 0.2 \pm 0.04$, see also the related
earlier calculation in Ref.~\cite{Bedaque:2010hr}.  
While the calculated chiral
extrapolations for $E_{\rm ^2H}$ are consistent with our earlier
analysis in \cite{Epelbaum:2002gb} as well as with the recent
phenomenological calculation of Ref.~\cite{Flambaum:2007mj},
unquenched lattice-QCD results of the NPLQCD Collaboration \cite{Beane:2009py}
seem to indicate an opposite trend with a stronger-bound deuteron at
large values of $m_q$. It is not clear at this stage whether there is
any contradiction since the lattice results are so far only available
at rather large pion masses with $M_\pi > 353.7$ MeV, see \cite{Detmold}.
Using the available lattice data in conjunction with the (presumably
unrealistic) assumptions 
of (i) perturbativeness of the $1\pi$-exchange potential in the
$^3S_1$-$^3D_1$ channel and (ii) validity of the chiral expansion for
NN scattering at
such large values of $M_\pi$ leads to a qualitatively different
dependence of $E_{\rm ^2H}$ on $m_q$ \cite{Chen:2010yt,Soto:2011tb}.    

Complementary to the studies based entirely on $\chi$EFT, there are   
interesting recent developments towards merging $\chi$EFT with dispersion
relations
\cite{Albaladejo:2011bu,Gasparyan:2012km}. 
The main idea of this approach is to take advantage of the known analytic
structure of the NN scattering amplitude $T(s)$, with $s$ being the
invariant mass of the NN system, which can be reconstructed from
the so-called generalized potential $U(s)$ by means of the (once
subtracted) dispersion relation  
\begin{equation}
 T(s\,)=U(s\,)+\, \int_{4m_N^2}^{\infty}\frac{ds'}{\pi}\,\frac{s-\mu_M^2}{s'-\mu_M^2}\,
\frac{T(s)\,\rho(s')\,T^*(s')}{s'-s-i\epsilon}\,,
\label{def-non-linear}
\end{equation}
where $\rho (s)$ is the phase-space function (see Ref.~\cite{Gasparyan:2012km} for the
exact definition). Further, $\mu_M$ denotes the matching point for which $T(
\mu_M) = U( \mu_M)$.  The generalized potential does not have the
right-hand elastic unitarity cut but still features left-hand cuts
associated with $t$-channel pion exchanges and short-range
mechanisms. The discontinuity across the first left-hand cut is
unambiguously given by the $1\pi$-exchange potential. The discontinuity
across the left-hand cut in the range from $s=4 m_N^2 - 4 M_\pi^2$
to $s=4 m_N^2 - 9 M_\pi^2$ is calculated in
Ref.~\cite{Gasparyan:2012km} at N$^2$LO using a manifestly covariant
version of ChPT. This relies on the assumption of the
validity of ChPT for the NN amplitude in some  region below
threshold. 
Extrapolating the
contributions from more distant left-hand cuts in $U(s)$ to the physical
region by means of a suitable conformal mapping
\cite{Gasparyan:2010xz,Danilkin:2010xd} or, more precisely, the Taylor
expansion in a conformal variable $\xi(s)$, we solved in
Ref.~\cite{Gasparyan:2012km} 
the partial-wave projected nonlinear equation (\ref{def-non-linear})
for $T(s)$ using the $N/D$ method.   Fixing the constants entering 
the expansion in $\xi(s)$ of the short-range part of $U(s)$ 
in $S$- and $P$-waves from NN phase shifts up to the energy
of $E_{\rm lab} = 100$ MeV, the energy dependence 
could be reasonably well described up to $E_{\rm lab} = 250$ MeV.   
We also observed good convergence of the chiral expansion for 
$U(s)$ when going from the order $Q^0$ to $Q^3$ (which supports the
assumption about perturbativeness of the amplitude below threshold)
and clear evidence for the $1\pi$- and $2\pi$-exchange cuts in NN
phase shifts.

\subsection{The three-nucleon force}
\label{3NFstatus}

Three-nucleon forces (3NF) represent an old but still very current 
topic in nuclear physics, see Refs.~\cite{KalantarNayestanaki:2011wz,Hammer:2012id} for recent review
articles. While effects of 3NFs in low-energy nuclear 
observables are expected to be considerably smaller than the ones of
the NN  force, see Eq.~(\ref{nuclforces}), their inclusion is
necessary at the level of accuracy of today's few- and
many-body {\it ab-initio} calculations, see
\cite{KalantarNayestanaki:2011wz,Hammer:2012id} and references
therein. In spite of decades of effort, the
structure of the 3NF is not properly described by the
available phenomenological models 
\cite{KalantarNayestanaki:2011wz}. Given the very rich spin-momentum
structure of the 3NF, scarcer database for
nucleon-deuteron (Nd)
scattering compared to the NN system and
relatively high computational cost of solving the Faddeev equations, further progress in this fields
requires substantial input from theory. This provides a strong motivation
to study the 3NF within $\chi$EFT. 

\begin{figure}[t]
\begin{minipage}{0.42\textwidth}
\vspace{-5.4cm}
\includegraphics[width=\textwidth,keepaspectratio,angle=0,clip]{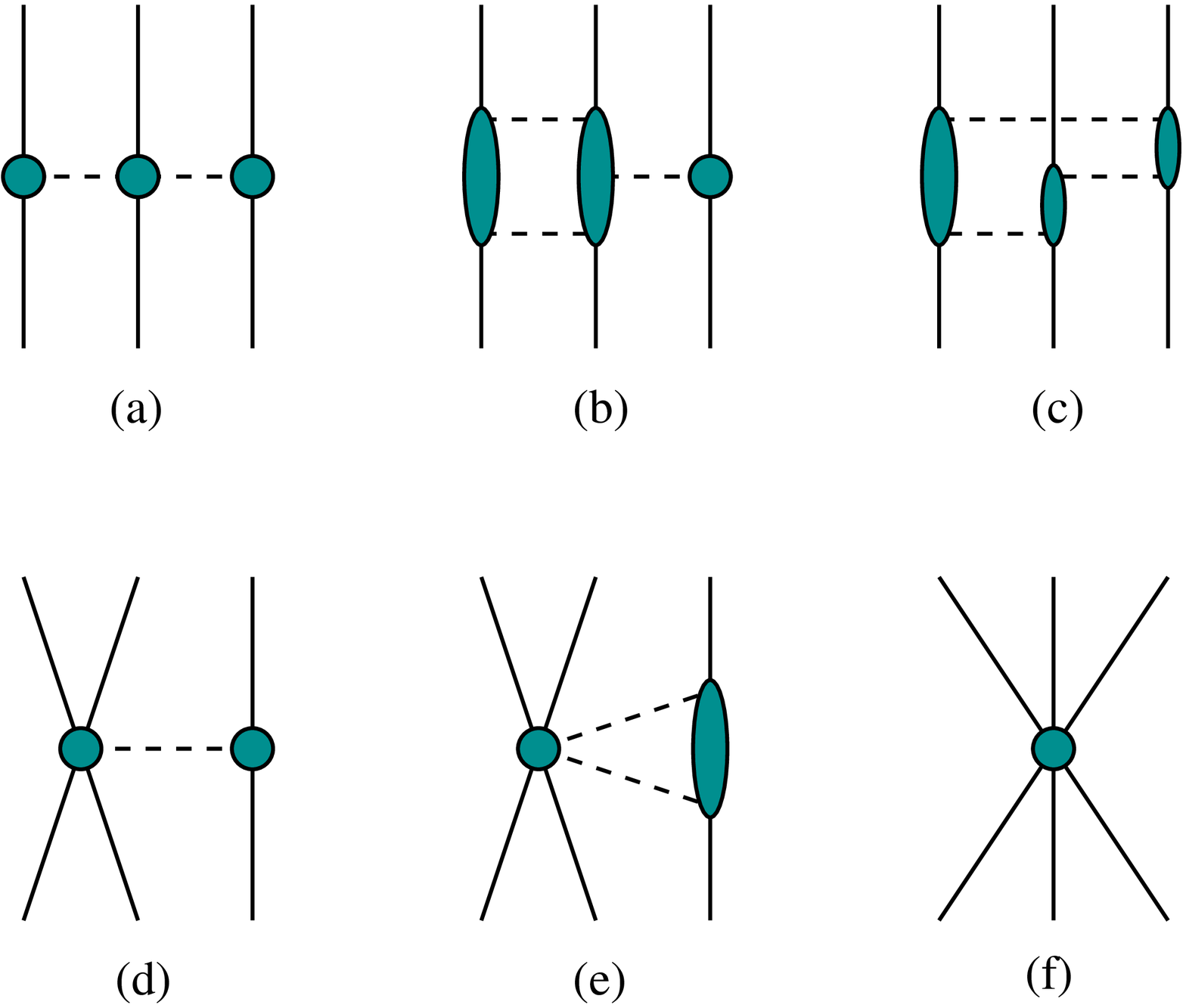}
\end{minipage}
\hfill
\includegraphics[width=0.48\textwidth,keepaspectratio,angle=0,clip]{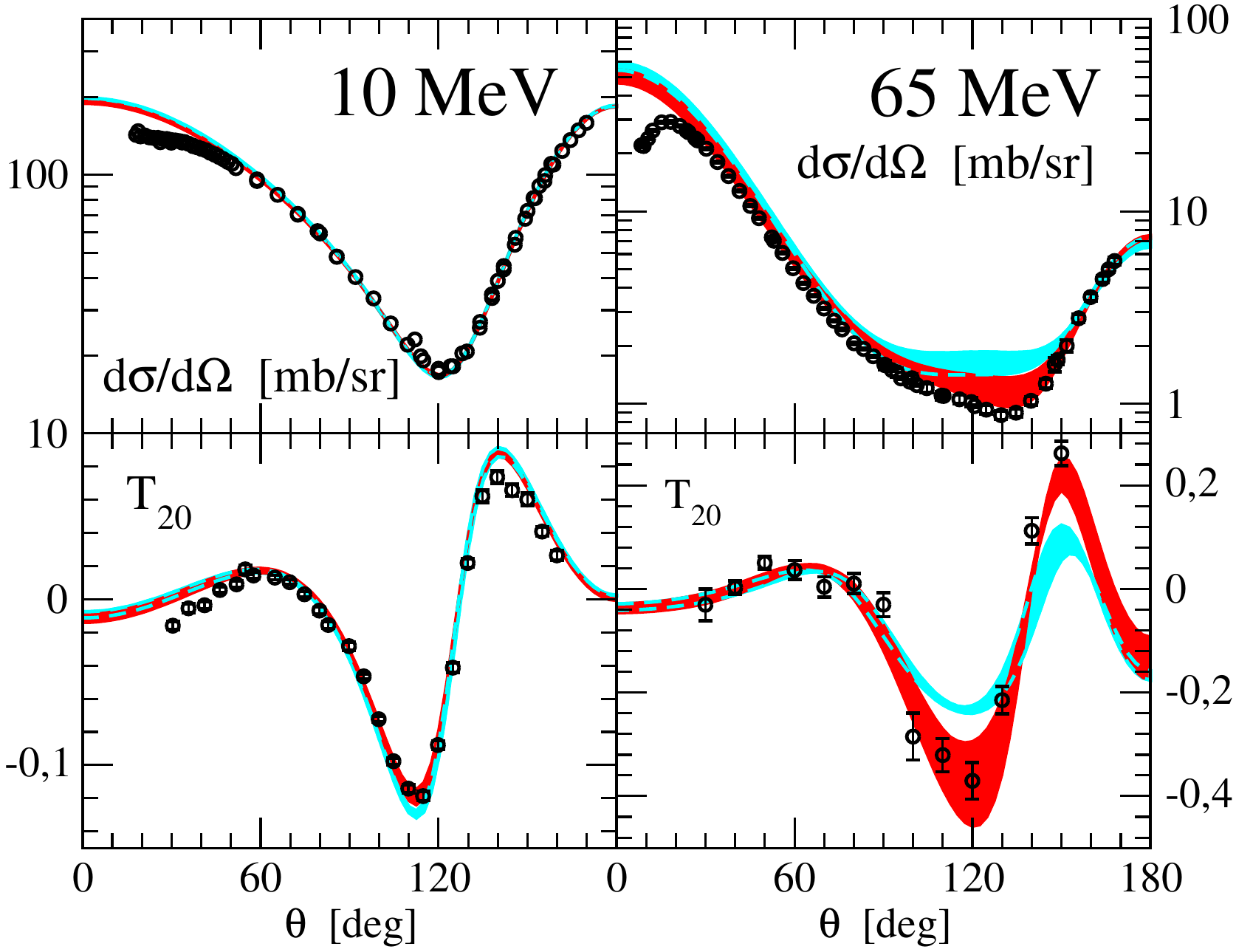}
    \caption{
          Left panel: Various topologies contributing to the 3NF up to and
          including N$^4$LO. 
Solid and dashed lines
         represent nucleons and pions while  
shaded blobs denote the corresponding amplitudes.  Right panel: Differential cross section and tensor analyzing
         power $T_{20}$  for elastic Nd scattering at
         $E_{\rm lab}^N=10$ and $65$ MeV. Cyan and red shaded bands
         correspond to NLO and N$^2$LO, respectively. 
\label{fig:3N} 
 }
\end{figure}

The first nonvanishing contributions to the 3NF 
emerge at N$^2$LO from tree-level diagrams in the left panel of
Fig.~\ref{fig:3N} corresponding to the
$2\pi$-exchange, $1\pi$-contact and pure contact graphs
(a), (d) and (f), respectively
\cite{vanKolck:1994yi,Epelbaum:2002vt}. 
The shorter-range terms emerging from 
diagrams (d) and (f) depend on one unknown LEC each which can be
determined from suitable few-nucleon observables, see
e.g.~\cite{Epelbaum:2008ga,Epelbaum:2012vx,KalantarNayestanaki:2011wz,Hammer:2012id}
and references therein. The long-range contribution (a) is driven by
the subleading $\pi \pi NN$ vertices proportional to the LECs
$c_{1,3,4}$ which are known from $\pi$N scattering. 
The resulting 3NF at N$^2$LO has been extensively explored in three-
and four-nucleon scattering calculations, see
\cite{KalantarNayestanaki:2011wz} and references therein.  One
finds a good description of low-energy Nd scattering
observables, see Fig.~\ref{fig:3N} for representative examples, except
for the well-known, long-standing puzzles such as 
the vector analyzing power in elastic Nd scattering (the so-called
$A_y$-puzzle) and  the cross section in the space-star
breakup configuration, see Ref.~\cite{KalantarNayestanaki:2011wz} for more
details. Promising results for low-energy four-nucleon
scattering observables based on the chiral 3NF, especially in
connection with the $A_y$-puzzle, are reported in
Ref.~\cite{Viviani:2010mf}.  While Nd scattering data at higher energies are
also well described, the theoretical uncertainty increases rapidly 
reflecting a similar pattern in the NN sector at this order in the
chiral expansion. This is visualized in the right panel of
Fig.~\ref{fig:3N}. 

Interesting results based on chiral nuclear forces were also obtained
by various groups in nuclear structure calculations showing, in particular, sensitivity to
the individual terms of the N$^2$LO  3NF, see \cite{Hammer:2012id}
for a review. All these findings clearly underline the need
to include corrections to the 3NF beyond the leading terms at N$^2$LO.

\subsection{Open challenges and ongoing work}

\subsubsection{Renormalization of the NN scattering amplitude with
  non-perturbative pions}

While the  long-range part of the nuclear force obeys the standard chiral power counting, the
relative importance of the short-range operators and the closely
related issue of renormalization of the Lippmann-Schwinger (LS) equation
are still under debate, see 
\cite{Nogga:2005hy,PavonValderrama:2005wv,Epelbaum:2006pt} 
and references therein for a sample of different points of view. 
The main problem is due to the fact that iterations of the truncated NN
potential within the LS equation generate contributions to the
amplitude beyond the order one is working at. These higher-order terms
generally involve ultraviolet (UV) divergences which cannot be
absorbed by counter terms (contact interactions) included in the
truncated potential so that one needs to employ a finite UV cutoff $\Lambda$ 
of the order of a natural hard scale, say $\Lambda \sim \Lambda_\chi
\sim M_\rho$ \cite{Lepage:1997cs}. All calculations
described in the previous sections have been carried out within such
an approach. Notice that it is \emph{not} legitimate  to 
employ $\Lambda \gg M_\rho$  (even if the limit 
$\Lambda \to \infty$ of the amplitude exists) 
unless \emph{all} UV divergences appearing 
in the iterations of the LS equation are subtracted
\cite{Epelbaum:2009sd}, see Ref.~\cite{Zeoli:2012bi} for an
illustration.   While
subleading and higher-order corrections to the potential do not \emph{have
to} be resummed in Weinberg's power counting scheme \cite{Weinberg:1990rz} and can be treated
perturbatively, it is easy to see that already the LS equation for the LO NN potential
\begin{equation}
\label{LO}
V_{\rm 2N}^{(0)} = - \frac{g_A^2}{4 F_\pi^2}  \fet \tau_1 \cdot
\fet \tau_2 \frac{\vec \sigma_1 \cdot \vec q \; \vec \sigma_2 \cdot
  \vec q}{\vec q \, ^2 + M_\pi^2} + C_S + C_T \vec \sigma_1 \cdot \vec \sigma_2,
\end{equation}
where $\vec \sigma_i$ ($\fet \tau_i$)  denote the Pauli
spin (isospin) matrices of a nucleon $i$ and $\vec q = \vec p \,
' - \vec p$ the nucleon momentum transfer,
is not renormalizable (in the above-mentioned sense).  In
\cite{Epelbaum:2012ua} we have shown that this unpleasant feature is
caused by the nonrelativistic expansion of the NN propagator
adopted in the LS equation. We have suggested a new framework based on
the manifestly Lorentz invariant effective Lagrangian.  
In this approach the LO amplitude is obtained by solving
the integral equation (first introduced in Ref.~\cite{kadyshevsky})
\begin{equation}
{T (
\vec p\,',\vec p )}{=}{ V _{\rm 2N }^{(0)} (
\vec p\,',\vec p ) - \frac{m_N^2}{2}\, \int \frac{d^3 k}{(2\,\pi)^3}}
{\frac{V _{\rm 2N}^{(0)} (
\vec p\,',\vec k ) \, T (
\vec k,\vec p )}{(k^2+m_N^2 )\, (
E-\sqrt{k^2+m_N^2}+i\,\epsilon)},}
\label{kad}
\end{equation}
where $E=\sqrt{p^2+m_N^2}$ denotes the center-of-mass energy of a single nucleon. 
Iterations of this equation generate only logarithmic
divergences which can be absorbed into a redefinition of 
$C_S$ and $C_T$, i.e. it is perturbatively
renormalizable. Consequently, the UV cutoff $\Lambda$ can be safely
removed by taking the limit $\Lambda \to \infty$.   Partial
wave projected equations corresponding to Eq.~(\ref{kad}) have unique
solutions except for the $^3P_0$ channel. The non-uniqueness of the
solution in this partial wave can be dealt with by resumming the
corresponding counter term, see \cite{Epelbaum:2012ua} for more
details. We further emphasize that the nucleon mass appearing in the
integrand in Eq.~(\ref{kad}) does not violate the power
counting after
renormalization is carried out
\cite{Epelbaum:2012ua,Epelbaum:2013ij}. We already applied this novel
scheme to NN scattering at LO. The resulting cutoff-independent
phase shifts and mixing angles are shown in
Fig.~\ref{fig:phases}. Given that the calculations are carried out at LO, the
agreement with the Nijmegen PWA is rather good. 

Since we do not rely on the nonrelativistic
expansion and do not attempt to integrate out the momentum scale $\sqrt{M_\pi
  m_N}$, our new scheme can be straightforwardly applied to study the
$m_q$ ($M_\pi$)
dependence of nuclear observables. At LO this is entirely driven by the
explicit $M_\pi$-dependence of the $1\pi$-exchange  potential in
Eq.~(\ref{LO}). Fig.~\ref{fig:mpidep} shows the predicted chiral
extrapolations of $E_{^2\rm H}$, $a_{1S0}^{-1}$ and $a_{3S1}^{-1}$. It is comforting to
see a good agreement with the N$^2$LO calculations of
Ref.~\cite{Berengut:2013nh} based on the formulation with a finite cutoff, see section
\ref{sec:NN}.  

In the future, these calculations should be extended to higher
orders by perturbatively including 
corrections
to the potential.  Recent studies
\cite{Valderrama:2011mv,Long:2011xw} within the
nonrelativistic 
framework indicate that such a perturbative treatment of the
$2\pi$-exchange might be phenomenologically successful.

\subsubsection{Three-nucleon force beyond N$^2$LO}

The first corrections to the 3NF emerge at N$^3$LO from all possible one-loop
diagrams of type (a)-(e) in Fig.~\ref{fig:3N} constructed from solely
the LO vertices. The resulting parameter-free
expressions can be found in
Refs.~\cite{Bernard:2007sp,Bernard:2011zr}, see also Ref.~\cite{Ishikawa:2007zz}.
An interesting feature of the N$^3$LO 3NF corrections is their
rather rich isospin-spin-momentum structure emerging primarily from
the ring topology (c) in Fig.~\ref{fig:3N}. This is in  contrast with the quite
restricted operator structure of the N$^2$LO 3NF. The new structures
in the 3NF might have impact on Nd scattering observables and
shed light on the existing puzzles. Numerical implementation of the
N$^3$LO 3NF corrections requires their partial wave
decomposition (PWD) which is a nontrivial
task. In Ref.~\cite{Golak:2009ri}, a new method to perform the PWD 
of any type of the 3NF by carrying out five-dimensional
angular integrations numerically was introduced. 
The PWD of the N$^3$LO  3NF using this new technique requires
substantial computational resources and is in
progress, see Ref.~\cite{Skibinski:2011vi} for some first (but still incomplete) results. 

\begin{figure}[t]
\begin{minipage}{0.42\textwidth}
\vspace{-6.74cm}
\includegraphics[width=\textwidth,keepaspectratio,angle=0,clip]{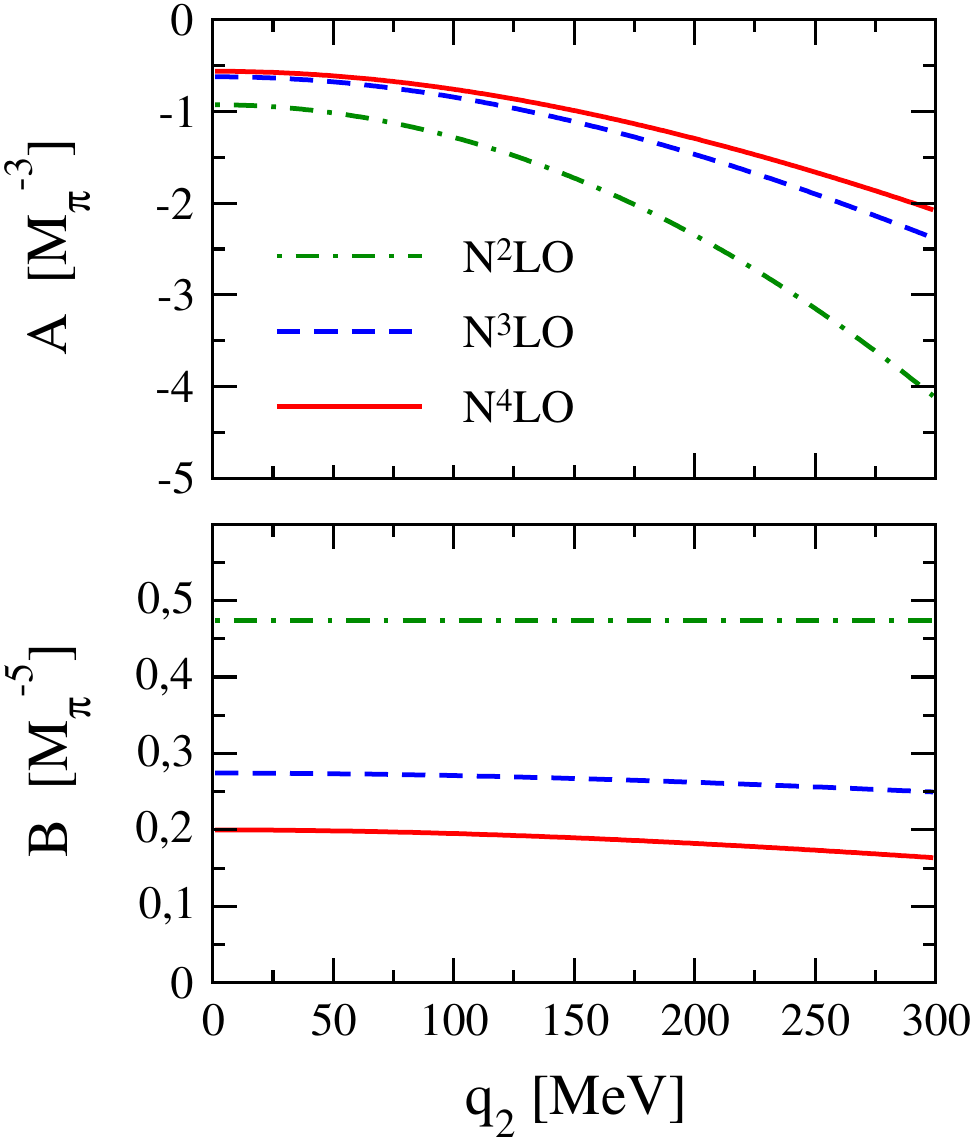}
\end{minipage}
\hfill
\includegraphics[width=0.50\textwidth,keepaspectratio,angle=0,clip]{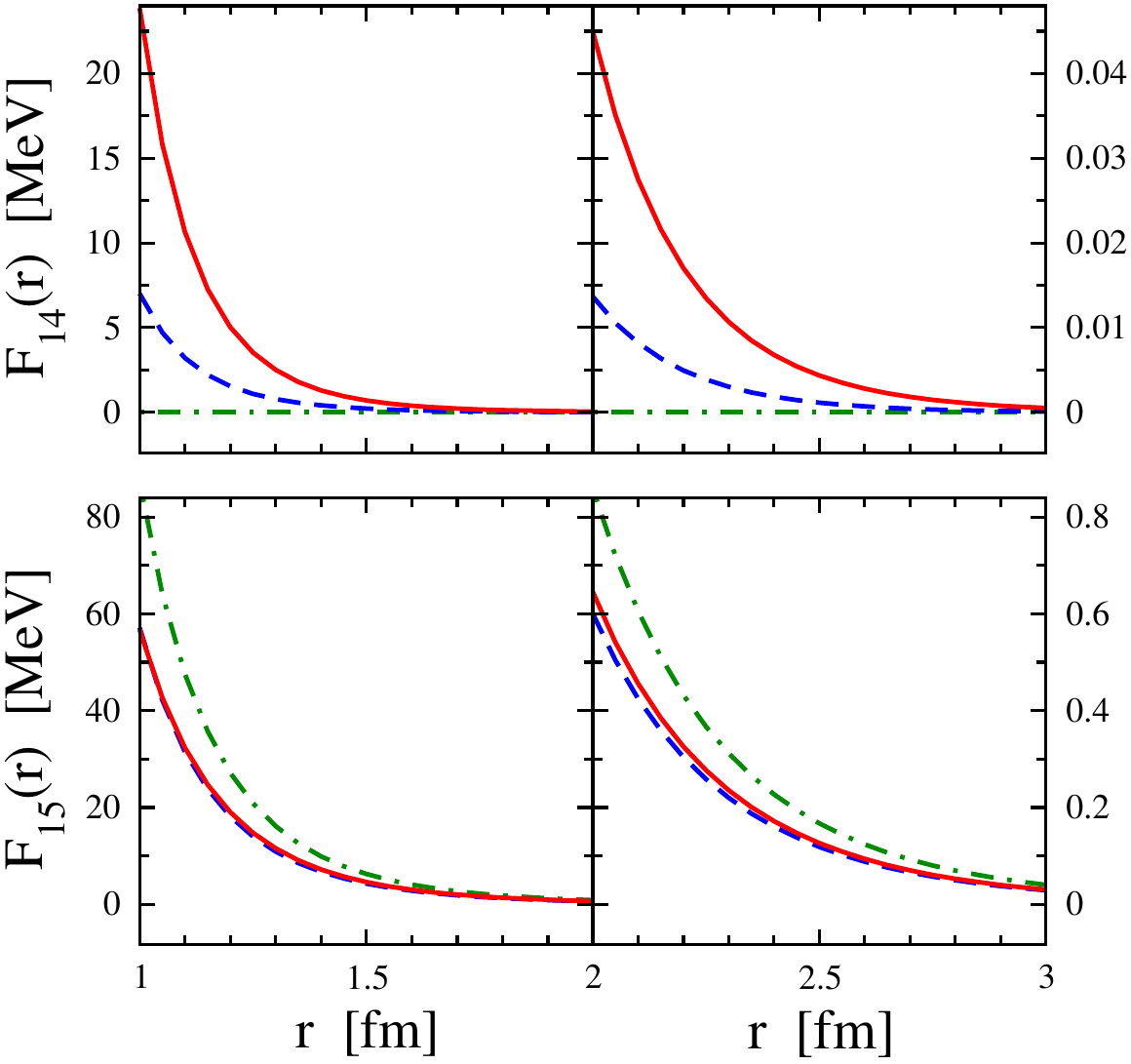}
    \caption{
          Left panel: Chiral expansion of the functions ${\cal A} (q_2)$ and ${\cal B} (q_2)$
entering the $2\pi$-exchange 3NF. Right panel: Chiral expansion of the functions ${\cal F}_i
      (r)$ generated by the long- and intermediate-range 3NF
      topologies up to N$^4$LO. Dashed-dotted, dashed and solid lines correspond to
      ${\cal F}_i^{(3)}$,   ${\cal F}_i^{(3)} + {\cal F}_i^{(4)}$ and
      ${\cal F}_i^{(3)} + {\cal F}_i^{(4)} + {\cal F}_i^{(5)}$,
      respectively. 
\label{fig:3NFexpansion} 
 }
\end{figure}

Meanwhile, one may ask whether the chiral expansion of the  3NF 
is already converged at N$^3$LO. Given the situation with the
$2\pi$-exchange NN potential, where the strongest 
contributions driven by $\Delta$ excitations appear at the subleading
order (in that case N$^2$LO), one may expect a similar
convergence pattern for the 3NF. This applies especially to new operator structures emerging
from the genuine loop topologies (b) and (c), whose chiral expansion 
starts at N$^3$LO rather than N$^2$LO. At this order, the resulting
contributions completely miss effects associated with intermediate
$\Delta$(1232) excitations. To clarify the situation 
it is, therefore, necessary to go to the
next-higher order N$^4$LO which corresponds for connected 3N diagrams
to the subleading one-loop order.  
First steps along these
lines were made recently in Ref.~\cite{Krebs:2012yv}, where the chiral expansion of
the longest-range, $2\pi$ exchange 3NF topology was extended to
N$^4$LO.  In the isospin and static limits, the $2\pi$ exchange 3NF in momentum space 
has the form
\begin{equation}
\label{2pi_general}
V^{2 \pi}_{\rm 3N} = \frac{
\vec \sigma_1 \cdot \vec q_1\,  \vec \sigma_3 \cdot
 \vec q_3}{[q_1^2 + M_\pi^2 ] \, [q_3^2 + M_\pi^2 ]}  
\Big( \fet
 \tau_1 \cdot \fet \tau_3  \, {\cal A}(q_2) + \fet \tau_1 \times  \fet
\tau_3 \cdot \fet \tau_2  \,   \vec q_1 \times  \vec q_3   \cdot \vec
\sigma_2 \, {\cal B}(q_2) \Big)   +  5\mbox{ perm.}\,.
\end{equation}
The quantities ${\cal A} (q_2)$ and 
${\cal B} (q_2)$ are scalar
functions whose explicit form is
computed within the chiral expansion. Notice that the leading
nonvanishing contributions to ${\cal A}$ and ${\cal B}$ at N$^2$LO
are governed by the LECs $c_i$ and already take into account effects
of the $\Delta$ isobar, see the discussion in section \ref{3NFstatus}.  One may
therefore expect a good convergence of the chiral expansion for these
quantities.  At the N$^4$LO ($Q^5$) level, 
the functions ${\cal A}$ and ${\cal B}$ 
depend on certain combinations of LECs from the order-$Q^2$,
$Q^3$ and $Q^4$ effective $\pi N$ Lagrangian. Their values were
determined in Ref.~\cite{Krebs:2012yv}  from $\pi N$ scattering
calculated within the same power counting scheme up to the subleading
one-loop order. One then indeed observes a good convergence of the chiral
expansion for 
${\cal A}$ and ${\cal B}$, see the left
panel of Fig.~\ref{fig:3NFexpansion}, which is
fully in line with the qualitative arguments given above.  

The situation with the
intermediate-range contributions emerging from the $2\pi$ - $1\pi$ and
ring diagrams (b) and (c) in Fig.~\ref{3NFstatus} is completely
different. Here, effects of the $\Delta$-isobar start showing up at
N$^4$LO leading to large corrections at this order 
\cite{Krebs:2013}. It is
more natural to address the convergence of the chiral expansion for
long-range 3N potentials in coordinate space. The general structure of
a local 3NF in coordinate space can be parameterized in terms of 22
scalar functions ${\cal F}_i(r_{12}, r_{23}, r_{31} )$  \cite{Krebs:2013}
\begin{equation}
V_{\rm 3N} = \sum_{i=1}^{22}{\cal G}_i(\vec\sigma_1,\vec \sigma_2,
\vec\sigma_3,\fet \tau_1,\fet \tau_2, \fet \tau_3,\hat r_{12}, \hat
r_{23}){\cal F}_i(r_{12}, r_{23}, r_{31} )+5\,{\rm permutations}, 
\label{strf22}
\end{equation}
where $\hat r_{ij} \equiv \vec r_{ij}/| \vec r_{ij |}$ and $\vec
r_{ij}= \vec r_i - \vec r_j$ denotes the position of nucleon $i$
with respect to nucleon $j$. The explicit form of the 22 linearly
independent operators ${\cal G}_i$ can be found in
Ref.~\cite{Krebs:2013}. The profile functions ${\cal F}_i $ have dimension of energy 
and can be interpreted as the potential energy between three static
nucleons projected onto the corresponding operator. They receive
contributions from the long-range and intermediate-range 3NF topologies and are
predicted (at long distances) by ChPT.
This
is visualized in the right panel of Fig.~\ref{fig:3NFexpansion}, where
we show the functions ${\cal F}_{14}$ and  ${\cal F}_{15}$, which
accompany the operators ${\cal G}_{14}=\fet \tau_2 \cdot \fet \tau_3 \, \hat r_{12}
\cdot \vec \sigma_1 \,  \hat r_{12}
\cdot \vec \sigma_2$ and  ${\cal G}_{15}=\fet \tau_1 \cdot \fet \tau_3 \, \hat r_{13}
\cdot \vec \sigma_1 \,  \hat r_{13}
\cdot \vec \sigma_3$, in the equilateral triangle configuration with  $r_{12}= r_{23}= r_{31} =r$.
The Fourier transform of Eq.~(\ref{2pi_general}) corresponding to the
longest-range, $2\pi$ topology gives rise to 10 out of 22 functions
${\cal F}_i $. They strongly dominate over the intermediate-range
contributions at distances $r\gtrsim 2$ fm, so that the chiral
expansion of the resulting ${\cal F}_i $'s shows good
convergence. The function ${\cal F}_{15}$ 
in Fig.~\ref{fig:3NFexpansion} may serve as a representative example. On
the other hand, those functions ${\cal F}_i$ which do not receive
contributions from $2\pi$-diagrams such as e.g.~${\cal F}_{14}$ are
dominated by the large N$^4$LO corrections to the $2\pi$-$1\pi$ and
ring graphs.  Notice that in agreement with the power counting, see Eq.~(\ref{nuclforces}), 3N
potentials are considerably weaker than the NN ones, 
cf.~Fig.~\ref{fig1} and the right panel of
Fig.~\ref{fig:3NFexpansion}.   More work is needed to clarify whether
phenomenologically important effects associated with intermediate 
$\Delta$-excitations are already properly represented at N$^4$LO. This can be
naturally addressed within the $\Delta$-full formulation of $\chi$EFT.
Work along  these lines is in progress. Last but not least, we
emphasize that subleading contributions to the 3N contact interactions
(diagram (f) in Fig.~\ref{fig:3N}), which also appear at N$^4$LO,
are worked out in Ref.~\cite{Girlanda:2011fh}.

\section{Precision few-nucleon physics: Recent examples}
\label{sec:applications}


Parallel to the developments in the strong sector,  there
have been several recent applications of $\chi$EFT to few-nucleon reactions with external
probes. Here we briefly discuss a few 
examples. 

\begin{figure}[t!]
\begin{center}
\begin{minipage}{0.42\textwidth}
\vskip -5.5 true cm
\includegraphics[width=\textwidth,keepaspectratio,angle=0,clip]{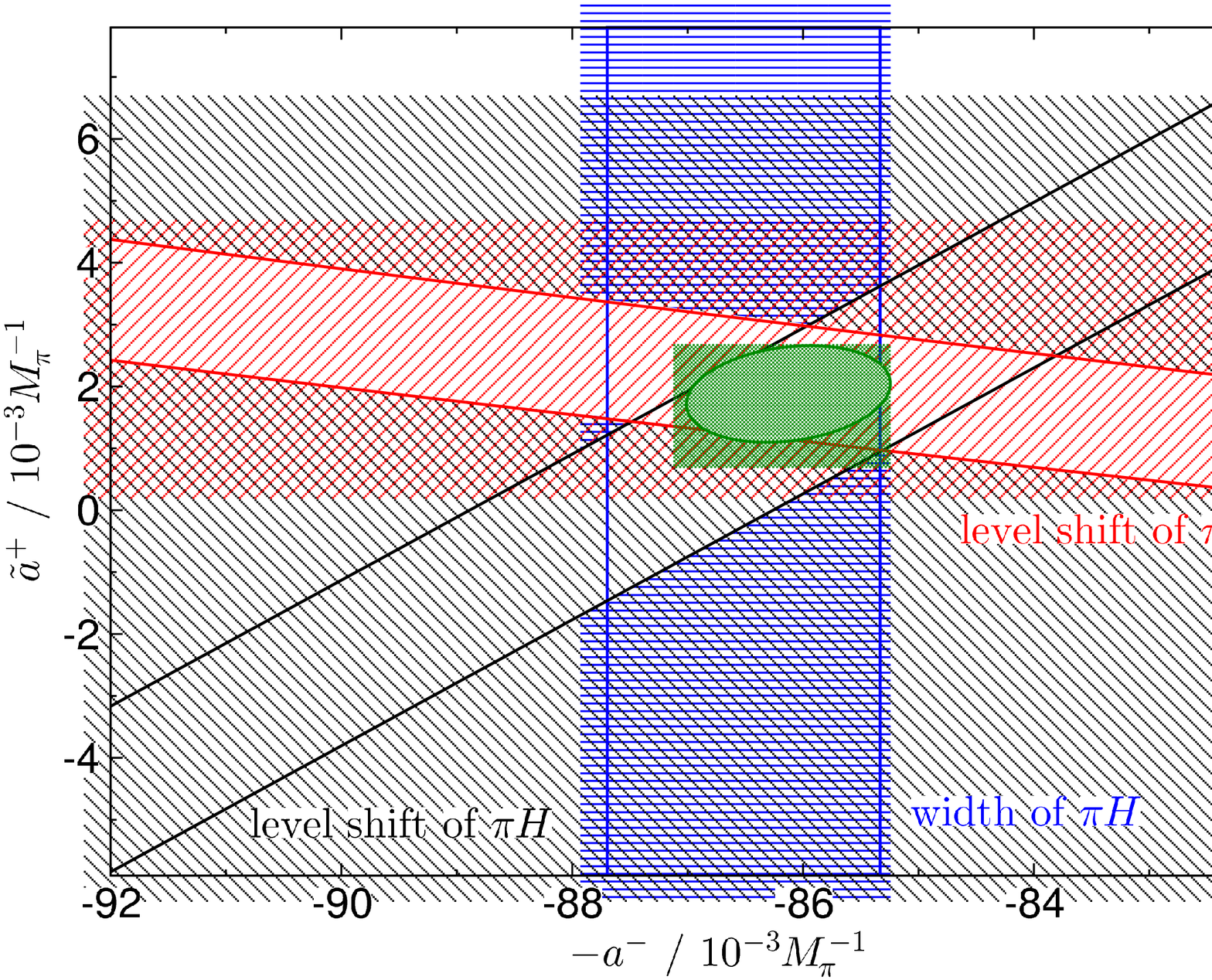}
\end{minipage}
\hfill
\includegraphics[width=0.48\textwidth,keepaspectratio,angle=0,clip]{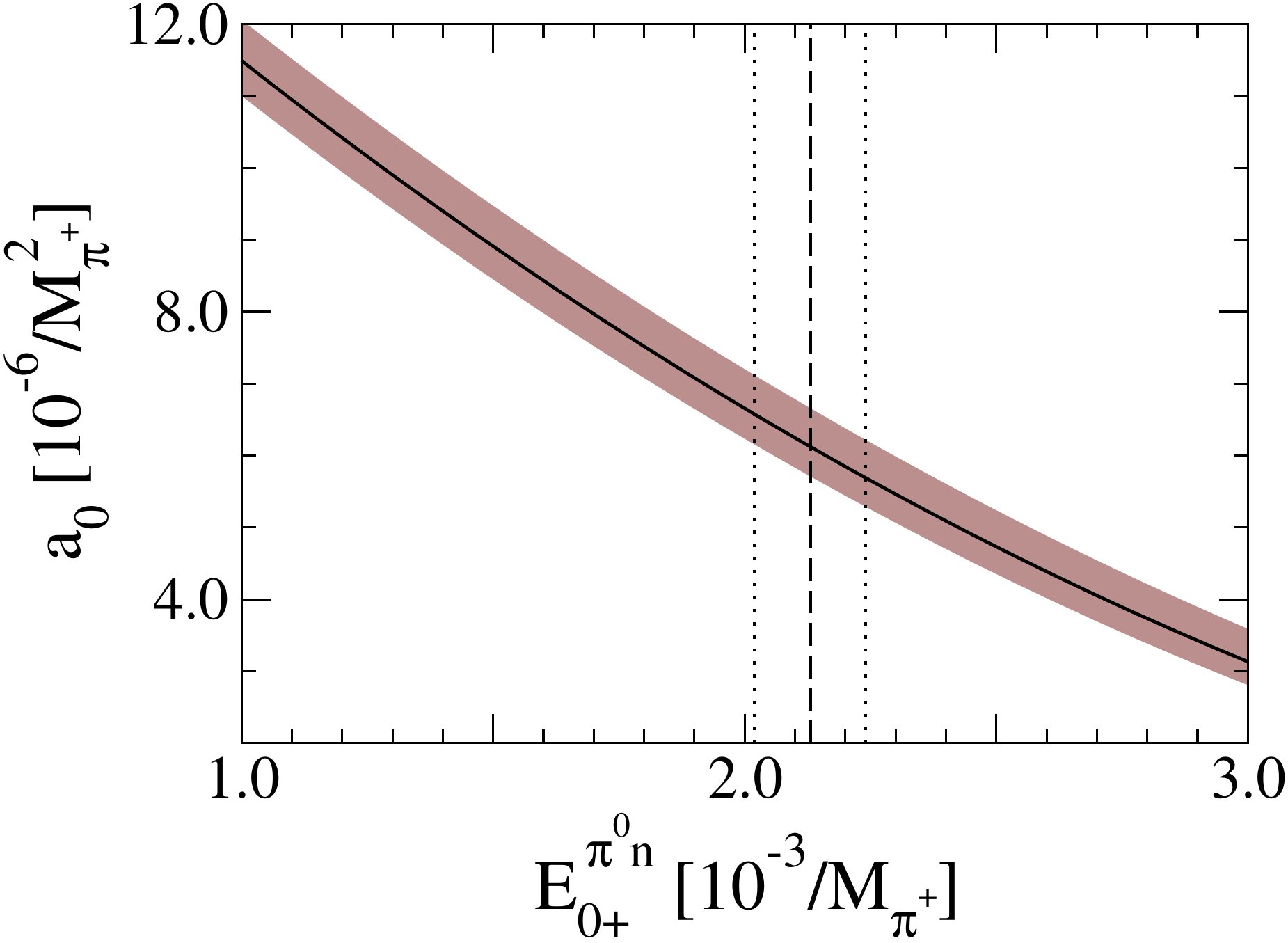}
\end{center}
\vspace{-0.5cm} 
    \caption{Left panel: Combined constraints on $\tilde a^+$ and
      $a^-$ from data on $\pi H$ and $\pi d$. Figure courtesy of Vadim
      Baru. Right panel:  sensitivity of $a_0$ for $^3$He in units
      of $10^{-6}/M_{\pi^+}^2$ to the neutron multipole
      $E_{0+}^{\pi^0 n}$. The vertical dashed line indicates  the ChPT
      prediction for $E_{0+}^{\pi^0 n}$ and the vertical dotted lines
      the estimated $5\%$ uncertainty.  The shaded band gives an
      indication of the theoretical uncertainty, see \cite{Lenkewitz:2011jd} for more
      detail.    
         \label{fig:ExtProbes} 
 }
\end{figure}

The first, classical example deals with low-energy pion-deuteron scattering
as a tool to extract the $\pi$N isoscalar and isovector scattering
lengths $a^+$ and $a^-$.  Direct extractions of $a^+$ 
from $\pi$N scattering  suffer from large
uncertainties, so that the important source of experimental
information on these fundamental observables 
comes nowadays from pionic atoms. Applying an improved Deser formula
\cite{Lyubovitskij:2000kk} to the 
experimentally 
measured shift and width of the $1s$
level of $\pi H$ $\epsilon_{1s} = (-7.120 \pm 0.012)$ eV and 
$\Gamma_{1s} = (0.823 \pm 0.019)$ eV  \cite{Gotta:2008zza} yields
constraints on the scattering lengths shown in
Fig.~\ref{fig:ExtProbes}.  
Here, $\tilde a^+$ denotes a particular
linear combination of $a^+$ and isospin-violating terms which is
accessible in pionic atoms, see \cite{Baru:2011bw}
for more details. The $\pi H$ data alone do not fix 
even the sign of $\tilde a^+$.  On the other hand, further constraints
can be obtained from the recently measured level shift of pionic
deuterium \cite{Strauch:2010vu} $\pi D$, $\epsilon_{1s}^D = (2.356 \pm 0.031)$ eV provided
one can reliably relate this observable to
$\pi N$ scattering lengths.  Such a theoretical analysis in
the framework of $\chi$EFT was recently accomplished in \cite{Baru:2011bw},
see also references therein for earlier calculations. Notice that at
this level of accuracy,  it was
necessary to carefully study isospin-breaking effects. 
It is comforting to see that the resulting constraints on $\tilde a^+$
and $a^-$ are consistent with those emerging from the $\pi H$
data. Moreover, 
combining $\pi H$ and $\pi D$ data leads
to the best-available determination of $\tilde a^+$  and $a^-$.    

Neutral pion photo- and electroproduction off light nuclei has also been
extensively explored in recent years, see Refs.~\cite{Beane:1997iv,Krebs:2004ir} and references
therein for earlier studies.  Here, one important motivation is to
test the ChPT prediction for the neutron amplitude,  $E_{0+}^{\pi^0 n}
= 2.13 \cdot 10^{-3}/M_{\pi^+}$ \cite{Bernard:2001gz}, which appears to be counterintuitive
given the ChPT value  for the proton $E_{0+}^{\pi^0 p}
= -1.16 \cdot 10^{-3}/M_{\pi^+}$ \cite{Bernard:2001gz,Bernard:2005dj} (which has been confirmed
experimentally). While the predicted value for $E_{0+}^{\pi^0 n}$ was
verified within $20\%$ by an experiment at SAL using deuteron target
\cite{B6_SALpid}, it is particularly promising to study $\pi^0$ production
off $^3$He which is known to be a good neutron target for
spin-dependent observables. Our recent theoretical calculation \cite{Lenkewitz:2011jd} at
subleading one-loop order of $\chi$EFT confirms these estimations. It is found
that the threshold S-wave cross section for $\pi^0$ photoproduction
off $^3$He, $ a_0 =( |\vec{k}_\gamma| / |\vec{q}_\pi| ) (
d\sigma / d\Omega )_{\vec{q}_\pi=0}$, is indeed sensitive to the
elementary neutron multipole $E_{0+}^{\pi^0 n}$, see the right panel
of Fig.~\ref{fig:ExtProbes}. Moreover, the
uncertainty associated with nuclear corrections appears to be very
small so that a reliable theoretical extraction of $E_{0+}^{\pi^0 n}$
from $a_0$ is possible. Using the above mentioned ChPT prediction for
$E_{0+}^{\pi^0 n}$, the predicted value of the $^3$He S-wave
multipole $E_{0+}$ is roughly consistent with the value deduced from
the old Saclay measurement of \cite{Argan:1987dm}.   

There is also considerable interest in electromagnetic few-nucleon
reactions.  In
the single-photon approximation, their theoretical description
requires knowledge of the electromagnetic current 
operator, which should be constructed consistently with the nuclear
forces. The derivation of the exchange currents in $\chi$EFT was first
addressed in the seminal paper by Park et al., \cite{Park:1995pn},
who, 
however, limited themselves to threshold kinematics.  
Recently, this work was extended by the JLab-Pisa \cite{Pastore:2009is} and Bochum-Bonn groups
\cite{Kolling:2009iq} to derive the exchange currents at
the one-loop level for general
kinematics suitable to study e.g.~electron
scattering off light nuclei at momentum transfer of the order of $\sim
M_\pi$. I do not discuss this topic any further due to the lack of
space but refer the reader to Refs.~\cite{Kolling:2012cs} and
\cite{Pastore:2012rp} for recent applications to the magnetic form
factor of the deuteron and magnetic moments of light nuclei.  
Finally, progress in the $\chi$EFT treatment of  Compton scattering off
light nuclei is summarized in Ref.~\cite{Griesshammer:2012we}. 

\section{Effective field theory on the lattice}
\label{sec:lattice}

Recently, a \emph{discretized} version of $\chi$EFT has been
developed and successively applied to compute the 
properties of few- and many-nucleon systems.  In this framework,
pions and nucleons are treated as point-like
particles on an Euclidean space-time lattice, and the path
integral is evaluated by Monte Carlo sampling
\cite{Muller:1999cp,Lee:2004si,Borasoy:2006qn,Lee:2008fa}. Using hadronic
DOFs allows one 
to probe large volumes and greater numbers of nucleons as compared to lattice
QCD. Clearly, the method is only applicable at low energies where
$\chi$EFT is expected to converge.  

The crucial object to calculate in our nuclear lattice simulations is the
correlation function for $A$ nucleons in the Euclidean
space-time, $Z_A(t)=\langle\Psi_A|\exp(-t H)|\Psi_A\rangle$, where 
$|\Psi_A\rangle$ denotes the Slater determinants for $A$ free 
nucleons, $H$ is the Hamiltonian of the system and $t$ the Euclidean
time. The correlation function can be most efficiently calculated
using the Hubbard-Stratonovich transformation to get rid of 
terms in the action quartic in the nucleon fields (at the expense of introducing
interactions with auxiliary fields) and
employing the hybrid Monte Carlo technique, see \cite{Lee:2008fa}
for more details.  Once $Z_A(t)$ is computed, the 
ground state energy of the $A$-nucleon system can be extracted from the 
asymptotic behavior of the correlation function for large $t$,
$E_A^0=-\lim_{t\rightarrow\infty} d(\ln Z_A)/dt$. 
In a similar way one can also obtain energies of low-lying excited
states and compute expectation values of normal ordered operators. 
%

\begin{figure}[t!]
\begin{center}
\begin{minipage}{0.57\textwidth}
\vspace{-4.7cm}
\begin{tabular}{|c||c|c|c|c|} \hline
& \hspace{.3cm} $0_{1}^{+}$ \hspace{.3cm}
& \hspace{.1cm} $2_{1}^{+}(E^{+})$ \hspace{.1cm} 
& \hspace{.3cm} $0_{2}^{+}$ \hspace{.3cm}
& $2_{2}^{+}(E^{+}) $ \\ \hline
LO 
& $-96(2)$ & $-94(2)$ & $-89(2)$ & $-88(2)$ \\
NLO 
& $-77(3)$ & $-74(3)$ & $-72(3)$ & $-70(3)$ \\
NNLO 
& $-92(3)$ & $-89(3)$ & $-85(3)$ & $-83(3)$ \\ \hline
Exp & $-92.16$ & $-87.72$ & $-84.51$ &
$-82(1)$
\\ \hline
\end{tabular}
\caption{
Left panel: Lattice results of Ref.~\cite{Epelbaum:2012qn} for the energies of low-lying even-parity
states of $^{12}$C compared to experimental values (in units of
MeV). Right panel: ``Survivability bands'' of carbon-oxigen based life
obtained from lattice simulations of Ref.~\cite{Epelbaum:2012iu} as explained in the text. 
\label{fig:Hoyle} 
 }
\end{minipage}
\hfill
\includegraphics[width=0.40\textwidth,keepaspectratio,angle=0,clip]{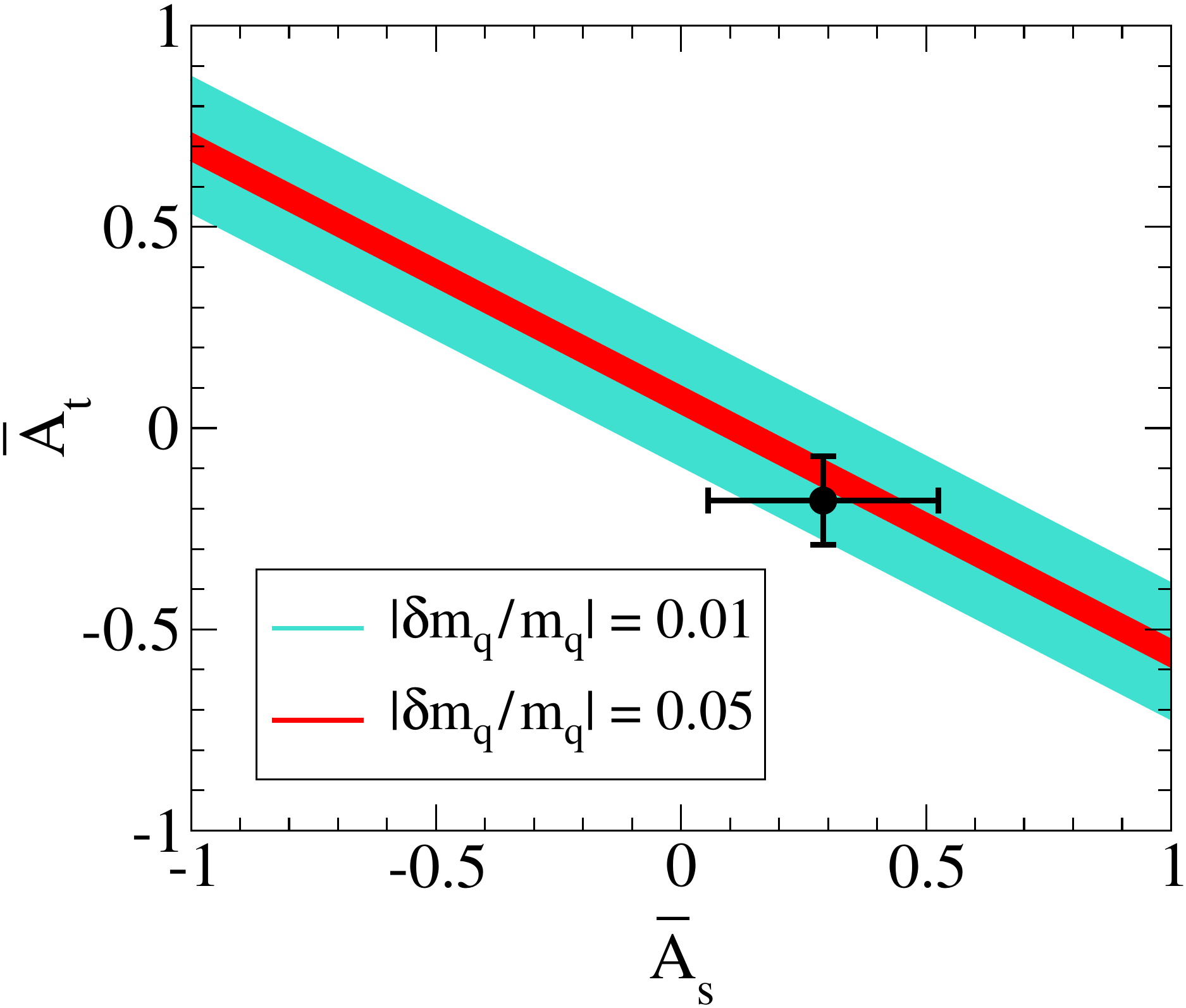}
\end{center}
\vspace{-0.6cm} 
\end{figure}

The most advanced studies within this framework are so far 
performed  at N$^2$LO in the chiral expansion.
The simulations are carried out using an (improved) LO action which incorporates the physics
of the $1\pi$-exchange and the LO NN contact
interactions. Higher-order corrections to the nuclear forces including
the Coulomb interaction and 3NFs are taken into account 
perturbatively, see \cite{Epelbaum:2010xt} for details. 
The resulting framework was recently applied to calculate the ground
state energies of $^4$He, $^8$Be and $^{12}$C as well as the low-lying
excitation energies in $^{12}$C including the the second spin-$0$ state, the famous Hoyle state 
\cite{Epelbaum:2011md,Epelbaum:2012qn}, see Fig.~\ref{fig:Hoyle}.      
All calculated energies are in a very good agreement with experiment. 
While the ground and spin-$2$ states of $^{12}$C have been also
calculated by other groups using different methods, the results for
the  Hoyle state are the first
\emph{ab initio} calculations. The Hoyle state is known to play a
crucial role for producing $^{12}$C, $^{16}$O and other elements
necessary for life via
the triple-alpha process in the red giant phase
of stars. The crucial quantity that controls the production rate 
is the energy $\epsilon$ of the Hoyle state relative to the
triple-alpha threshold which is experimentally known  to be $\epsilon = 397.47
(18)$ keV. Changing $\epsilon$ by an amount of $\pm 100$ keV 
results in a strong reduction of the formation of $^{12}$C and $^{16}$O in
the universe making the emergence of carbon-based life impossible. 
It is, therefore, very interesting to investigate how this
seemingly fine-tuned quantity depends on the fundamental constants of
nature such as $m_q$. We have studied the sensitivity of
$\epsilon$ to variations of $m_q$ within nuclear lattice simulations in
Ref.~\cite{Epelbaum:2012iu}. Fig.~\ref{fig:Hoyle} shows the
survivability bands of carbon-based life under $1\%$ and $5\%$ changes
of $m_q$. Here, $\bar A_{s,t} \equiv (\partial a^{-1}_{1S0,\, 3S1}/\partial
M_\pi)_{M_\pi^{\rm phys}}$ denote the slope of the inverse NN S-wave scattering lengths as
functions of the pion mass. These quantities can, in principle, be
computed in lattice-QCD. The data point in the right panel of
Fig.~\ref{fig:Hoyle} corresponds to the recent N$^2$LO results of
Ref.~\cite{Berengut:2013nh} for chiral extrapolations of
$a^{-1}_{1S0,\, 3S1}$  shown in Fig.~\ref{fig:mpidep}. These findings 
suggest that the formation of carbon and oxygen in our universe would survive
a $\sim 2\%$ change in the light quark mass.  


\end{document}